\documentclass[
aps,%
12pt,%
final,%
notitlepage,%
oneside,%
onecolumn,%
nobibnotes,%
nofootinbib,
superscriptaddress,%
noshowpacs,%
centertags]%
{revtex4} \RequirePackage{graphicx}

\def\refitem#1{\relax}

\begin{document}
\title{On possibility to observe chiral phase transition in separate fragments of dense baryon matter}

\author{\firstname{B.F.} \surname{Kostenko}\footnote{e-mail: bkostenko@jinr.ru}}
\affiliation{Joint Institute for Nuclear Research, Dubna, Russia}

\author{\firstname{J.} \surname{Pribi\v{s}}}
\affiliation{Technical University, Ko\v{s}ice,  Slovakia}

\begin{abstract}

Density fluctuations of intranuclear matter suffering collisions
with  projectile particles are capable to turn into multiquark
clusters with chiral symmetry restored. Theoretical analysis of
these processes requires an additional taking account of finite size
effects in the region of the chiral phase transition. From the
experimental point of view, this method of observation of the chiral
phase transition has its inherent advantages due to a relatively
moderate number of secondary particles to be registered.

\end{abstract}

\maketitle

\section{Introduction}
Studies of the two- and three-nucleon short range correlations
\cite{CLAS} afford us an opportunity to use the dense few-nucleon
correlated systems  of this type (SRC) as targets, which correspond
to small fragments of nuclear matter in the dynamically broken
chiral symmetry states. Collisions of SRC with bombarding particles
can initiate the chiral phase transition ending in the creation of
multibaryon (MB). Thus, its observation would be a direct evidence
of the chiral condensate disappearance in the interaction area.
Separation of a MB masses from the secondary particle background is
possible, if the excitation energy of MB is small enough. We suggest
to use the cumulative particle as a trigger for registration of MB
decay products, since its appearance is a signature of "deep
cooling" of the MB system.

\section{Kinematic analysis of MB production in cumulative processes}

Let us consider a cumulative particle, e.g. $\pi$-meson, outgoing
under angle $\theta$ with respect to momentum $\mathbf{p}_0$ of the
projectile proton. The law of conservation of energy-momentum gives:
\begin{equation}\label{1}
 E_p +M = E_{\pi} + (p_*^2 +M_*^2)^{1/2},
\end{equation}
where $M$ is a value of mass of SRC, undergoing the collision, $M_*$
is a value of mass of a total system in the final state aside from
the cumulative meson, $\mathbf{p}_*$ is momentum of the system,
$$
p_*=(p_{\pi}^2 -2p_{\pi} \cos \theta  + p_0^2)^{1/2}.
$$
Relation (1) may be considered as a function
\begin{equation}\label{2}
M_* = M_* (M),
\end{equation}
since all the values it contains are known from the cumulative
experiment arrangement. We use of $M_*(M)$  for estimation of
production possibility of $(n+1)$-baryon from $n$-nucleon SRC in the
cumulative process of this type. An interesting property of function
(2), merely mathematical one,  is its ability to demonstrate the
inhibition of the process for values of $M$ large enough. Therefore,
the possibility of production is estimated at the minimal mass $m_n$
of n-nucleon SRC. As  is known, SRC have a continuous mass spectrum
calculated at least for $2N$ and $3N$ SRC. In particular, $2N$ SRC
dominate the nuclear wave function at $k_{min} \ge $ 300 MeV
\cite{CLAS}. Relying on this, and taking into account an approximate
proportionality of SRC masses  to their baryon numbers, $n$, we
accept
$$
m_n \ge n,
$$
where $n$ is taken in GeV/c$^2$ units.  MB masses, $M_n$, are
estimated here in the quark bag model framework  \cite{LukTit}: $
M_3=3.62, \qquad M_4=4.76, \qquad M_5=6.07 $ GeV/c$^2$. The
criterion of possibility of the transition  $n$-nucleon SRC to
$(n+1)$-MB is
$$
M_* \ge M_{n+1}.
$$
Excitation spectrum of MB may reach values up to $ E_{ex}=(M_* -
M_{n+1})c^2. $

In Tables I and II estimations of possible yield of MB in cumulative
$\pi$-meson production from nuclear targets Be and Al for two angles
$\theta= 119^\circ $ and $\theta= 97^\circ $ are shown. The first
column encloses momenta of the registered cumulative mesons, the
fourth column contains baryon numbers of resonances which production
are possible. The invariant cross-sections, $f=A^{-1} E
d\sigma/d^3p$, for corresponding cumulative processes in
mb$\cdot$GeV$^{-2}$$\cdot$c$^3$$\cdot$sr$^{-1}$$\cdot$nucleon$^{-1}$
\cite{ITEF} are shown in the second and third columns. Here they
only represent an upper bound for production cross-section of
corresponding MB. One can see that values of $\theta$ and momentum
of cumulative $\pi$-meson should be maximal for observation of the
chiral phase transition in fragments of nuclear matter large enough.

In the parton model framework, high momentum cumulative mesons hold
SRC quarks, which had large momenta pointed at the opposite
direction as compared to momentum of projectile. Escape of these
quarks from the system results in deep cooling of MB. This
circumstance is favorable for discrimination of MB mass from the
secondary particle background, which production is stimulated by
increasing phase volume of the system.

\section{Finite size effects}

Research of chiral phase transition in small fragments of nuclear
matter calls for a careful analysis of finite size effects
stipulated by shell structure of quark bag, its surface energy, and
Coulomb forces at short distances. Quark bag models predict nearly
constant value of $M_n/n$ for $n=$ 3, 4, 5, corresponding to its
bulk value (see, e.g., \cite{LukTit, Farhi, de Swart}). A slight, on
2-3 $\%$ accuracy level, deviation from this value is caused by
shell effects. In accordance with \cite{Farhi}, surface tension
coefficient for quark bag is about (70 MeV)$^3 \approx$ 8,8
MeV/fm$^2.$ This gives for $M_3$ a correction about 2 -- 3 $\%$ for
radius of tribaryon estimated as radius of flucton, $R \approx 0.8$
fm \cite{LukTit}. Independent consideration of Casimir energy, which
includes the contribution of surface tension energy, gives the same
estimation within the bounds of the chiral bag model \cite{Vepstas}.
A compound (n$+$1)-baryon system, consisting of the projectile
proton and n-baryon SRC, can acquire an additional mass increase due
to the Coulomb repulsion of the charge of the projectile and a
charge of SRC. This gives a correction to $M_3$ on 0.13 $\%$ level
or less. Thus, our estimations indicate that the finite size effects
have no sufficient effect on the chiral phase transition occurrence.

\begin{acknowledgments} We thank A.V. Efremov, V.A. Nikitin,  A.V.
Stavinskiy and M.Z. Yuriev for critical remarks and illuminating
discussions. J.P. is grateful for the hospitality extended to him
during visits at JINR.
\end{acknowledgments}

\newpage

\begin{table}
\caption{\label{tab1}\small An upper bounds for production
cross-section of MB by 10.14 GeV protons irradiating nuclear targets
of Be and Al. Cumulative pion is registered at laboratory angle
$\theta= 119^{\circ} $.}
\begin{center}
\begin{tabular}
{|l |c | c| c| c| l|} \hline
$P_{\pi},$ ÃýÂ/c & f, Be & f, Al & B \\
\hline
0.873 & 1.65 $10^{-4} $ & 4.61 $10^{-4}$ &       \\
0.979 & 2.47 $10^{-5} $ & 8.62 $10^{-5}$ & 3     \\
1.077 & 3.72 $10^{-6} $ & 1.72 $10^{-5}$ & 3     \\
1.293 & 6.23 $10^{-8} $ & 3.56 $10^{-7}$ & 3     \\
1.402 & 8.21 $10^{-9} $ & 5.32 $10^{-8}$ & 3, 4  \\
1.512 & 7.94 $10^{-10}$ & 4.95 $10^{-9}$ & 3, 4  \\
1.619 &                 & 1.03 $10^{-9}$ & $\;$3, 4, 5 $\;$\\
\hline
\end{tabular}
\end{center}
\end{table}

\begin{table}
\caption{\label{tab1}\small The same experiment as in Table I but
$\theta= 97^{\circ} $.}
\begin{center}
\begin{tabular}
{|l |c | c| c| c| l|} \hline
$P_{\pi},$ ÃýÂ/c & f, Be & f, Al & B \\
\hline
1.192 & 1.95 $10^{-5} $ & 7.09 $10^{-5}$ &      \\
1.370 & 1.20 $10^{-6} $ & 6.34 $10^{-6}$ & 3    \\
1.523 & 9.36 $10^{-8} $ & 6.37 $10^{-7}$ & 3    \\
1.635 & 1.40 $10^{-8} $ & 1.26 $10^{-7}$ & $\;$ 3, 4 $\;$ \\
1.790 & 1.21 $10^{-9} $ & 1.42 $10^{-8}$ & $\;$ 3, 4 $\;$ \\
\hline
\end{tabular}
\end{center}
\end{table}

\end{document}